\documentclass[doublecol,figures]{epl2}
\usepackage{CJK}
\usepackage[caption=false]{subfig}     
\graphicspath{ {./images/} }
\usepackage{ifpdf}\ifpdf 
    \usepackage{hyperref}
    \PassOptionsToPackage{unicode}{hyperref}
\else
    \providecommand{\texorpdfstring}[2]{#1}
    \usepackage{hyperref}
\fi
\usepackage{siunitx,amsmath,amssymb}
\usepackage{cases,booktabs}
\graphicspath{ {./images/} }
\title{Time-averaged approach to the dewetting problem at evaporation}
\author{Xiaolong Zhang (张晓龙) \and Vadim S. Nikolayev\thanks{E-mail: \email{vadim.nikolayev@cea.fr}}}
\shortauthor{X. Zhang and V. S. Nikolayev}
\institute{Service de Physique de l'Etat Condens\'e, CEA Paris-Saclay, CNRS, Universit\'e Paris-Saclay, 91191 Gif-sur-Yvette Cedex, France}
\abstract{
Dewetting of liquid films on solid surfaces in the presence of evaporation is a common phenomenon and has been studied by many researchers.
The previous numerical approach has revealed that evaporation accelerates the dewetting speed of the triple contact line and established correlations between the dewetting speed and the surface wettability and superheating. However, such a numerical calculation is time- and resource-consuming.
We examine dewetting physics and propose a time-averaged approach based on the multiscale theory. The new approach averages the dewetting process over time and consists of only several algebraic equations, making the problem easier to solve.
It can produce time-averaged values of essential quantities, such as the dewetting speed and contact angle as a function of superheating, which agrees with the previous numerical results. This simple approach is valuable for many applications, such as modeling pulsating heat pipes and describing the microlayer dynamics under growing vapor bubbles in nucleate boiling.}

\begin{document}
\begin{CJK*}{UTF8}{gbsn}
\maketitle
\end{CJK*}

\section{Introduction}
\begin{figure}
\onefigure[width=\columnwidth]{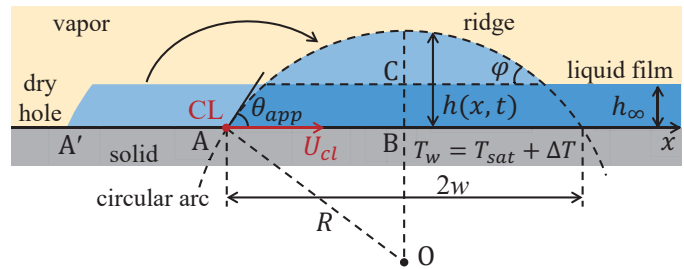}
  \caption{Dewetting of a liquid film: the ridge collects liquid as the contact line (CL) recedes.}\label{fig:DewettingRidge}
\end{figure}

Dewetting of liquid films is a common phenomenon, which can be observed on a daily basis and used in numerous industrial applications. We recapitulate the classical dewetting theory. Consider an ideally smooth and homogeneous solid surface initially covered by a liquid film of viscous fluid. Such a film is metastable under partial wetting conditions; once a dry hole is created in the film (the contact line appears), the dry area will grow spontaneously, because capillary forces drive the motion. As the dry hole enlarges, a liquid ridge forms ahead of the expanding contact line by collecting the liquid that previously covered the solid surface. Due to the high viscous friction in the thin film, the liquid in the ridge cannot re-enter back the film. Therefore, the ridge grows at the rate of liquid accumulation during the dry area expansion. This problem can be described in two dimensions, illustrated in fig.~\ref{fig:DewettingRidge}. Understanding the hydrodynamics of the contact line (CL) that demarcates the boundary between the wetted and dry solid surface parts is fundamental to the dewetting problem.

Until recently, the dewetting studies \cite{BW,Snoeijer10} concerned the adiabatic conditions (in the absence of net heat and mass transfer between liquid, solid and ambient gas). The work of \cite{JFM22} investigated dewetting in the presence of phase change over the liquid film (evaporation/condensation). The co-occurrence of the two processes is frequently encountered in studies of heat exchange systems, such as liquid film dynamics in pulsating heat pipes \cite{ATE21} and the drying of liquid microlayer beneath vapor bubbles in nucleate boiling \cite{Bures21a}. In these cases, the liquid film is surrounded by the pure vapor of the same fluid. The numerical results of \cite{JFM22} indicated that evaporation accelerates the dewetting; the contact line recedes faster when the solid substrate is heated. This Letter is structured as follows. First, we will briefly summarize the problem in \cite{JFM22} because these results will help to understand the multiscale analysis discussed next. Based on it new (``time-averaged'') approach will be finally introduced.

\section{2D lubrication problem}

\begin{figure}
  \onefigure[width=7cm]{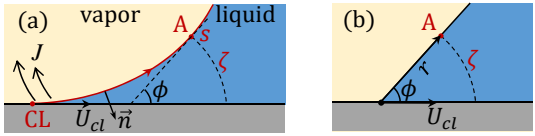}
  \caption{Sketches of the curved interface (a) and the straight liquid wedge (b) corresponding to the point A.}\label{fig:wedge}
\end{figure}

The problem is described in \cite{JFM22} in the 2D cross-section of the liquid ridge shown in fig.~\ref{fig:DewettingRidge}. It uses the ``one-sided'' formulation, where the vapor-side hydrodynamic stress and the heat flux into the vapor at the interface are neglected compared to those on the liquid side. Therefore, the vapor pressure $p_v$ is assumed to be spatially homogeneous. The solid substrate is highly conductive and thus isothermal. Its temperature $T_w=T_{sat}+\Delta T$ can differ from the saturation temperature $T_{sat}=T_{sat}(p_v)$, which induces liquid evaporation or condensation for positive or negative $\Delta T$, respectively. We focus on the evaporation case $\Delta T>0$, the most important for applications.

As the interface slope increases considerably in the presence of evaporation, the conventional lubrication theory, valid for slopes below $\sim 30^\circ$, becomes insufficient. The generalized lubrication theory \cite{ChestersB,Mathieu_thes,IPHT14} is used to describe the thin film flow and free liquid-vapor interface of local height $h(x)$. The theory uses the parametric interface description in terms of the curvilinear coordinate $s$ that runs along the interface (fig.~\ref{fig:wedge}a), with $s=0$ at the CL. Therefore, the following geometrical relations hold:
\refstepcounter{equation}\label{eq:param}
\begin{equation}
{\partial h}/{\partial s}=\sin \phi,\quad {\partial x}/{\partial s}=\cos \phi,
\tag{\theequation a,b}
\end{equation}
where $\phi$ is the local interface slope. The major convenience of this parametrization is the simplicity of rigorous expression for curvature $K$,
\begin{equation}\label{K}
K=\frac{\partial\phi}{\partial s}=\frac{1}{\cos\phi}\frac{\partial^2h}{\partial s^2}.
\end{equation}

Consider an interfacial point A (fig.~\ref{fig:wedge}a). A straight wedge can be formulated by the tangent line and the solid surface with the opening angle equal to the local interface slope $\phi$. The length of the intercepted arc $\zeta$ is expressed as $\phi r$, where $r$ is the radius of the straight wedge (fig.~\ref{fig:wedge}b) and is related to $h$ through $h=r\sin\phi$. Therefore, $\zeta=h\phi/\sin\phi$.

The generalized lubrication theory approximates the liquid pressure $p_l$ at A  (fig.~\ref{fig:wedge}a) by the pressure created by the flow in the corresponding straight liquid wedge with the opening angle $\phi$, cf. fig.~\ref{fig:wedge}b.

The interfacial pressure jump
\begin{equation}\label{eq:pJumpPr(s)}
\Delta p \equiv p_v-p_l=\sigma K -{J^2}( \rho_v^{-1} - \rho_l^{-1} ),
\end{equation}
accounts for the vapor recoil effect, where $K$ can be expressed with eq.~\eqref{K} and $J$ is the local mass flux assumed positive at evaporation.

Because the liquid films are thin, heat conduction is the main energy exchange mechanism, which can be taken as stationary due to their low thermal inertia. The liquid temperature is assumed to vary linearly along the arc $\zeta$ from $T_w$ to $T^i$, as suggested by the rigorous thermal analysis of straight wedges where the heat flow is radial \cite{Anderson1994}. The heat flux supplied to the vapor-liquid interface is spent to vaporize the liquid. Therefore, the energy balance at the interface reads
\begin{equation}\label{eq:J(s)}
  J = {k(T_w-T^i)}/({\zeta {\cal L}}),
\end{equation}
where $\cal {L}$ is the latent heat. The interfacial temperature $T^i$ is impacted by the Kelvin, vapor recoil, and molecular-kinetic effects:
\begin{equation}\label{eq:TintK-pr-Ri}
T^i = T_{sat}[ 1 + {\Delta p}/({{\cal L}\rho_l})  + {J^2}( \rho_v^{-2} - \rho_l^{-2})/({2{\cal L}}) ] +  R^i J {\cal L},
\end{equation}
where
\begin{equation}
\label{eq:Ri}
    R^i = \frac{T_{sat} \sqrt{2\pi R_v T_{sat}} (\rho_l-\rho_v) }{2 {\cal L}^2 \rho_l \rho_v }.
\end{equation}
is the kinetic interfacial thermal resistance. Here $R_v$ is the specific gas constant.

The governing equation
\begin{multline}\label{eq:GEA}
\underbrace{\frac{\partial h }{\partial t}\cos\phi}_A + \frac{\partial }{\partial s}\Bigg\{ \frac{1}{\mu G(\phi )}\Bigg[ \underbrace{\frac{\zeta}{2}( \zeta + 2 l_s )\frac{\partial \sigma }{\partial s}}_{B1}+\\ \underbrace{\frac{\zeta^2}{3}(\zeta+3l_s)\frac{\partial\Delta p}{\partial s}}_{B2} \Bigg] \underbrace{- U_{cl}\zeta \frac{F(\phi)}{G(\phi)}}_C \Bigg\} =  \underbrace{- \frac{J}{\rho_l}}_D,
\end{multline}
is written in the frame of reference of the moving CL, which is receding at speed $U_{cl}$. Here, $\mu$ and $\sigma$ are the liquid viscosity and surface tension, respectively; $l_s$ is the hydrodynamic slip length.

Several other effects counterbalance the capillarity (term $B2$). Since the interface temperature $T^i$ is not constant, cf. eq.~\eqref{eq:TintK-pr-Ri}, the flow  due to variation of surface tension is accounted for by the term $B1$,
\begin{equation}\label{eq:MarEff}
  {\partial \sigma }/{\partial s} \simeq  - \gamma {\partial T^i}/{\partial s},
\end{equation}
where $\gamma=-\partial\sigma/\partial T$ is generally positive. The term $C$ reflects the flow induced by the CL motion. The functions \cite{ChestersB,Mathieu_thes}
\begin{equation}\label{eq:F}
F(\phi)=\frac{2\phi^2}{3}\frac{\sin{\phi}}{\phi-\sin\phi\cos\phi},
\end{equation}
and
\begin{equation}\label{eq:G}
G(\phi)=\frac{\phi^3}{3}\frac{4}{\sin\phi\cos\phi-\phi\cos2\phi},
\end{equation}
are the correction factors. With $F(\phi\to 0)=1$ and $G(\phi\to 0)=1$, this formulation reduces to the conventional lubrication theory.

The flow induced by the phase change corresponds to the term $D$. Finally, the term $A$ describes the temporal ridge growth.

The governing equation \eqref{eq:GEA} can be solved for $s\in[0, s_f]$, with a point $s_f$ far from the ridge, where the interface is flat of thickness $h_\infty$. At $(s_f, h_\infty)$, the boundary condition is zero spatial derivatives $\phi=0$. Due to evaporation, $h_\infty$ decreases with time,
\begin{equation}\label{eq:hfe}
  h_\infty=h_0\sqrt{1-{t}/{t_d}},
\end{equation}
where $h_0$ is the initial ($t=0$) film thickness and
\begin{equation}\label{td}
t_d={{\cal L}\rho_l h_0^2}/({2k\Delta T})
\end{equation}
is the time of complete film drying.

At CL ($s=0$), the geometry implies
\begin{align}
& h= 0, \label{eq:BC1}\\
&{{\partial  h }}/{\partial s}= \sin\theta_{micro}, \label{eq:BC2}
\end{align}
where $\theta_{micro}$ is the microscopic contact angle controlling the wetting conditions. The set of lubrication equations (\ref{eq:pJumpPr(s)}, \ref{eq:GEA}) is of fourth order; one more boundary condition is needed. It expresses the solution regularity at CL \cite{PF10}, in particular, the pressure finiteness:
\begin{equation}\label{eq:BC3}
\left. {{\partial \Delta p}}/{\partial  s} \right|_{s \to 0}= 0.
\end{equation}

This form is convenient \cite{JFM21} to provide good numerical stability. Another equation determines the CL speed $U_{cl}$ (which is a part of the problem). The relation between $U_{cl}$ and the evaporation flux at CL \cite{JFM22}
\begin{equation}\label{eq:Jcl}
J(s\to 0)=\frac{U_{cl}F(\theta_{micro})}{\dfrac{G(\theta_{micro})}{\theta_{micro}\rho_l}+\dfrac{l_s{\cal L}\theta_{micro}}{\mu k}\gamma}.
\end{equation}

\begin{figure}
\centering
\subfloat[Liquid ridge growth.]
 {\includegraphics[width=6cm,clip]{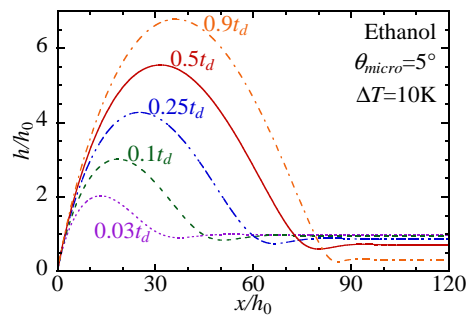}\label{fig:ethanolhxDT10}}\\
\subfloat[The CL receding deceleration.]
 {\includegraphics[width=6.5cm]{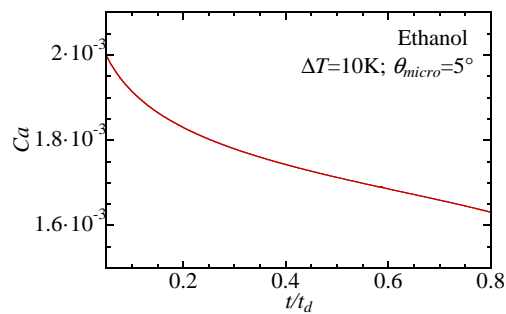}\label{fig:EthanolCa_t}}
\caption{Numerical results showing the evolution of dewetting ridge for ethanol at $p_v=\SI{23.2}{kPa}$; $\Delta T=10$K, $\theta_{micro}=5^{\circ}$, $l_s=\SI{10}{nm}$, and $h_0=\SI{50}{\micro m}$.}\label{fig:ethanol}
\end{figure}

A numerical result is illustrated in fig.~\ref{fig:ethanol}, which includes the interface profiles at selected time moments (in the CL frame reference) fig.~\ref{fig:ethanolhxDT10}, and the corresponding $U_{cl}$ dependence of time (expressed with the capillary number $Ca=\mu U_{cl}/\sigma$), fig.~\ref{fig:EthanolCa_t}.

Numerically solving eq.~\eqref{eq:GEA} is a delicate issue because it is a non-linear and non-stationary partial differential equation. Producing the results in fig.~\ref{fig:ethanol} took dozens of hours of calculation time on a regular PC.

\section{Multiscale analysis}

\begin{figure}[htp]
  \onefigure[width=6cm]{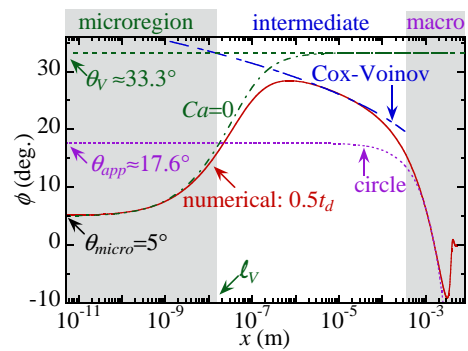}
  \caption{Spatial variation of the local interface slope showing different asymptotic regions (calculated for $t=0.5t_d$ and the same parameters as fig.~\ref{fig:ethanol}). }\label{fig:multiEthanol}
\end{figure}

\begin{figure}[htp]
  \onefigure[width=6.5cm]{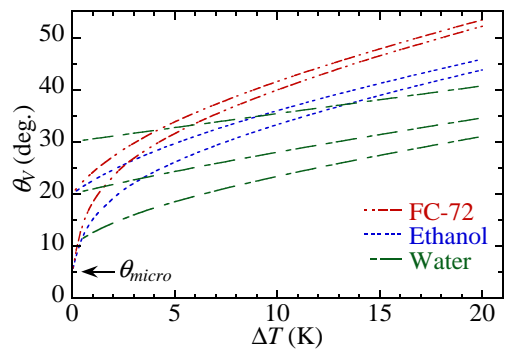}
  \caption{Variation of the Voinov angle $\theta_V$ with $\Delta T$ for several $\theta_{micro}=\theta_V(\Delta T=0)$ and three fluids: FC-72 ($\theta_{micro}=5^\circ,\,20^\circ$) at $p_v=\SI{46.7}{kPa}$, ethanol ($\theta_{micro}=5^\circ,\,20^\circ$) at $p_v=\SI{23.2}{kPa}$, and water ($\theta_{micro}=10^\circ,\,20^\circ,\,30^\circ$) at $p_v=\SI{101}{kPa}$ \cite{JFM22}.}\label{fig:thetaV_DT}
\end{figure}

Our new approach is based on the multiscale analysis of dewetting problem. To explain it, we plot in fig.~\ref{fig:multiEthanol} the interface slope $\phi(x)$ at $t=0.5t_d$ obtained from the numerical result (fig.~\ref{fig:ethanol}). Three regions can be identified \cite{PRE13movingCL,SWEP22}. In the microregion (the distance from the CL $x\lesssim \SI{100}{nm}$), phase change dominates the flow that is described by the balance of the terms $B1$, $B2$ and $D$ of eq.~\eqref{eq:GEA}. The CL motion term ($C$) and slow ridge growth ($A$) can both be neglected \cite{PRE13movingCL}. The microregion problem is thus steady, and the corresponding curve is identified for this reason as $Ca=0$ in fig.~\ref{fig:multiEthanol}. The microregion problem has been investigated by many researchers (see \cite{SWEP22} for a review). To solve the microregion problem, one applies the zero-pressure jump condition far from CL,
\begin{equation}\label{eq:p=0}
  \left.\Delta p\right|_{s \to \infty}= 0.
\end{equation}
The other three boundary conditions are (\ref{eq:BC1}-\ref{eq:BC3}). At $x\to\infty$, $\phi$ saturates to the value $\theta_V$ that is called hereafter the Voinov angle; in the case of fig.~\ref{fig:multiEthanol}, $\theta_V\simeq 33.3^\circ$;  $\theta_V(\Delta T=0)=\theta_{micro}$.  A particularity of the problem statement \cite{EuLet12} presented here is its applicability to the partial wetting case. Note that the solution results in $\theta_V$ as a function of the local superheating (that at CL, cf. fig.~\ref{fig:thetaV_DT}) and wetting properties.
Note that $\theta_V=\theta_{micro}$ at $\Delta T=0$.

Above the microregion, there is an ``intermediate'' region, within which the CL motion governs fluid flows (fig.~\ref{fig:multiEthanol}). The nanoscale effects: the hydrodynamic slip, vapor recoil, and Marangoni effects are nonessential here \cite{PRE13movingCL}. The interface is thus controlled by the balance of the terms $C$ and $B2$  (with $l_s\to 0$) in eq.~\eqref{eq:GEA}. It is the classical dynamic CL problem studied by Voinov and Cox, and has an asymptotic solution
\begin{equation}\label{eq:Cox-Voinov}
  \phi^3 = \theta^3_V-9Ca\ln ({x}/{\ell_V}),
\end{equation}
where the Voinov length $\ell_V$ can be found by fitting eq.~\eqref{eq:Cox-Voinov} (blue dashed curve in fig.~\ref{fig:multiEthanol}) to the full numerical solution \cite{PRE13movingCL}. In the case of fig.~\ref{fig:multiEthanol}, $\ell_V\simeq\SI{13}{nm}$. $\theta_V$ appears  in the intermediate region solution as an integration constant defined from the condition $ \phi(Ca\to 0) = \theta_V$.

Finally, at the macroscopic (millimetric) scale, the ridge growth is described by the balance of the terms $A$ and $B2$ in eq.~\eqref{eq:GEA}. Its profile is defined by the capillary forces only and can thus be fitted by a circular arc (violet dotted curve in fig.~\ref{fig:multiEthanol}). The limit $x\to 0$ of this latter solution results in the apparent contact angle $\theta_{app}$, which is an apparent slope at the CL seen from the macroscopic viewpoint: $\theta_{app}\simeq 17^\circ$ in this case.

A multiscale approach has been applied \cite{Snoeijer10} to solve the conventional adiabatic dewetting case $\Delta T=0$ (where  $\theta_V\equiv\theta_{micro}$ as mentioned above) by analytical asymptotic methods. They resulted in two expressions giving $\theta_{app}$ and $Ca$,
\begin{equation}\label{asymp}
\begin{split}
  &\theta^3_{app} =\theta^3_V-9Ca\ln [{2w}/({e\ell_V})],\\
  &Ca=\frac{\theta^3_V}{9}\left[ \ln \left(  \frac{4a}{e^2} Ca^{1/3} \frac{w^2}{\ell_V h_\infty} \right) \right] ^{-1},
\end{split}
\end{equation}
where $a\simeq 1.094$, $e=2.71\dots$ is the Euler number, and $w$ is the ridge half-width (fig.~\ref{fig:DewettingRidge}). The second asymptotic expression is obtained in the limit of large times, where the ridge growth is slow so the ridge is quasi-steady.

\section{Time-averaged model}
During the ridge growth, all the quantities, $Ca, w$ and $h_\infty$ (considering phase change), are not constant. $Ca$ slowly decreases with time, cf. fig.~\ref{fig:EthanolCa_t}; $w$ grows as the ridge collecting more liquid, and $h_\infty$ decreases because of evaporation, cf. eq.~\eqref{eq:hfe}. The variation in $\ell_V$ turns out to be small during the film evaporation \cite{JFM22}.

\begin{figure}
\centering
\subfloat[]{\includegraphics[height=4cm]{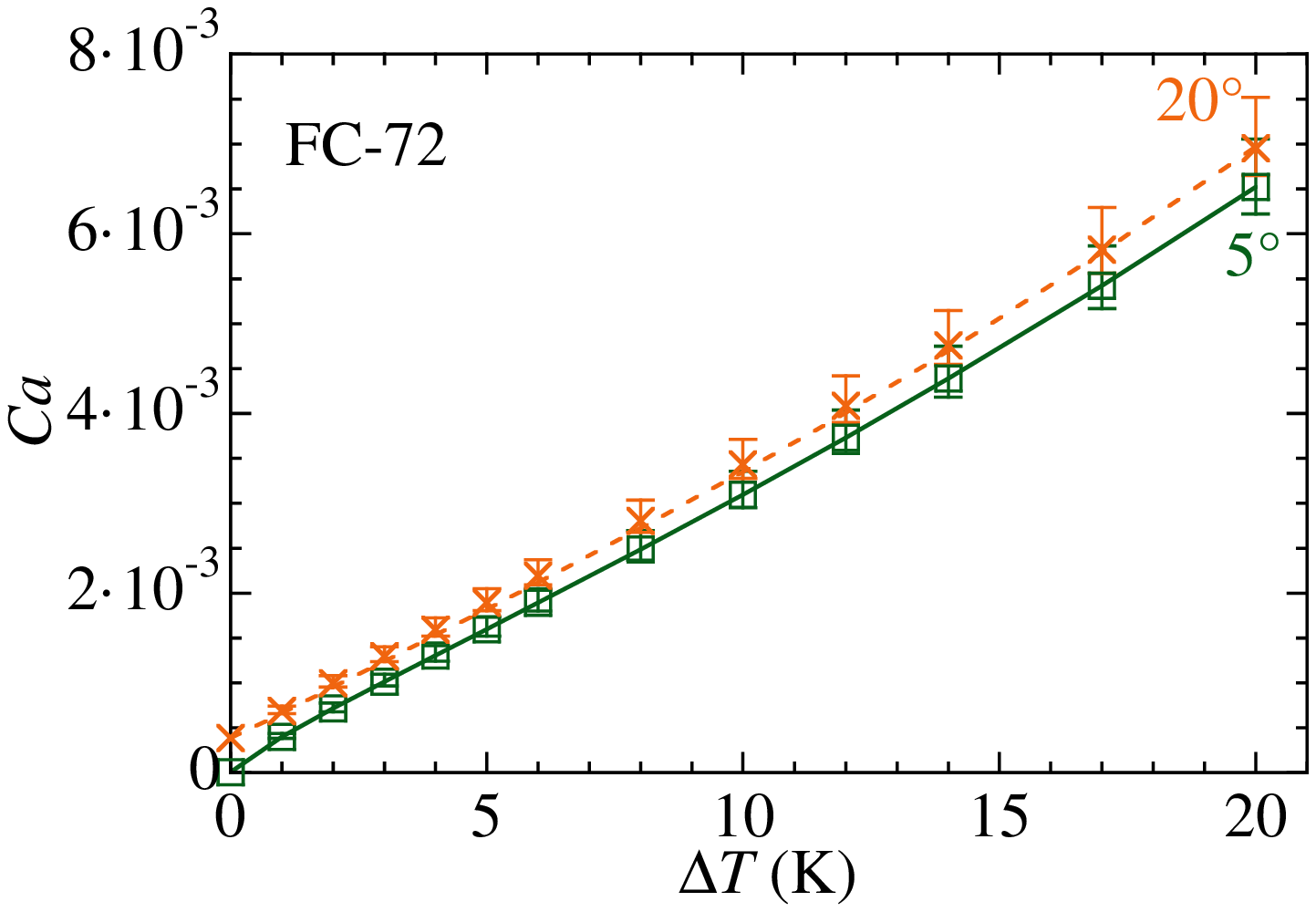}\label{fig:FC72CaDT}}\\
\subfloat[]{\includegraphics[height=4cm]{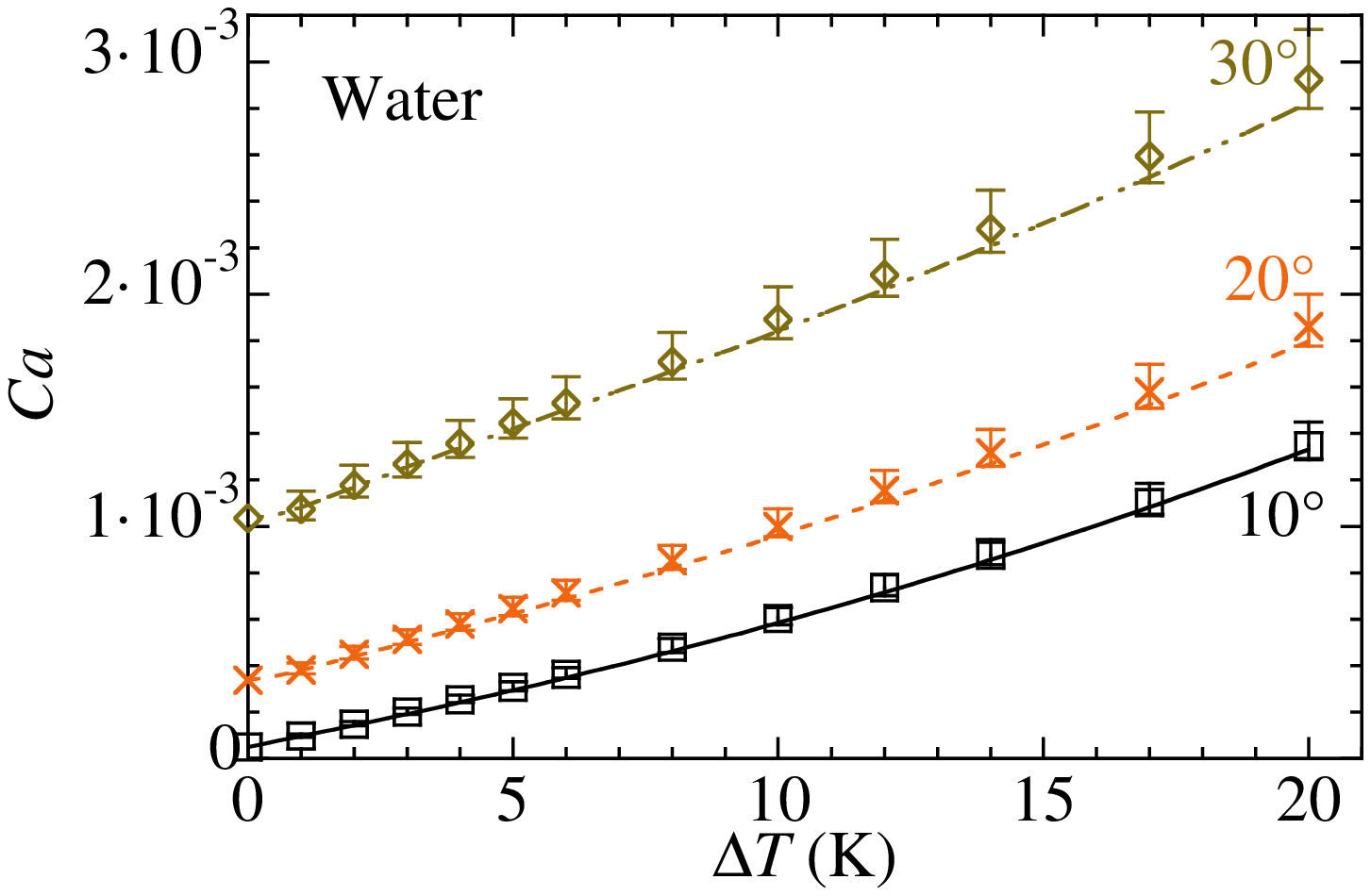}\label{fig:WaterCaDT}}\\
\subfloat[]{\includegraphics[height=4cm]{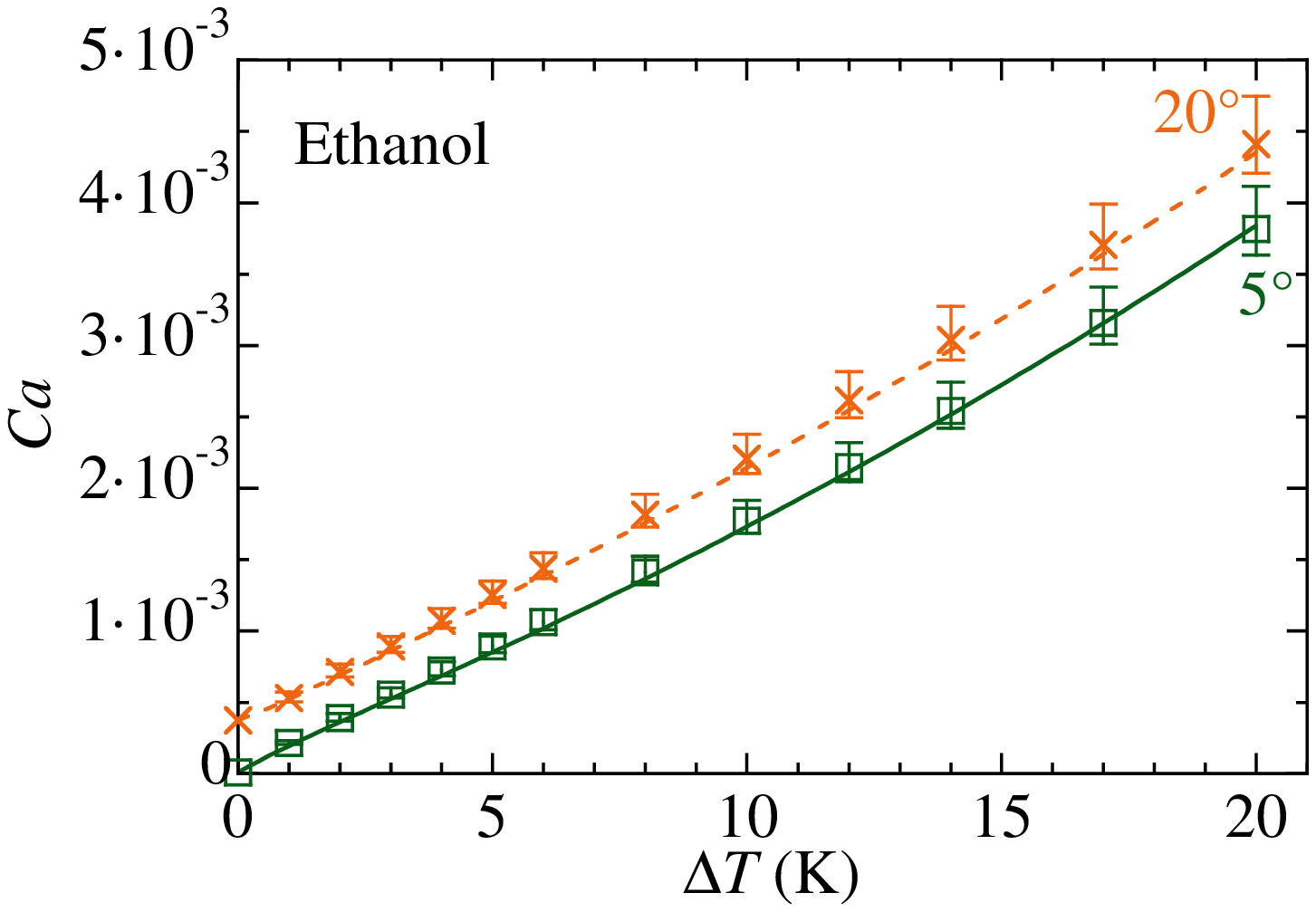}\label{fig:EthanolCaDT}}
  \caption{Comparison of numerical (characters with bars) and time-averaged model (lines) results of $Ca(\Delta T)$ for several $\theta_{micro}$ and different fluids.}\label{fig:CaDT}
\end{figure}

\begin{figure}
\centering
\subfloat[]{\includegraphics[width=5cm, clip]{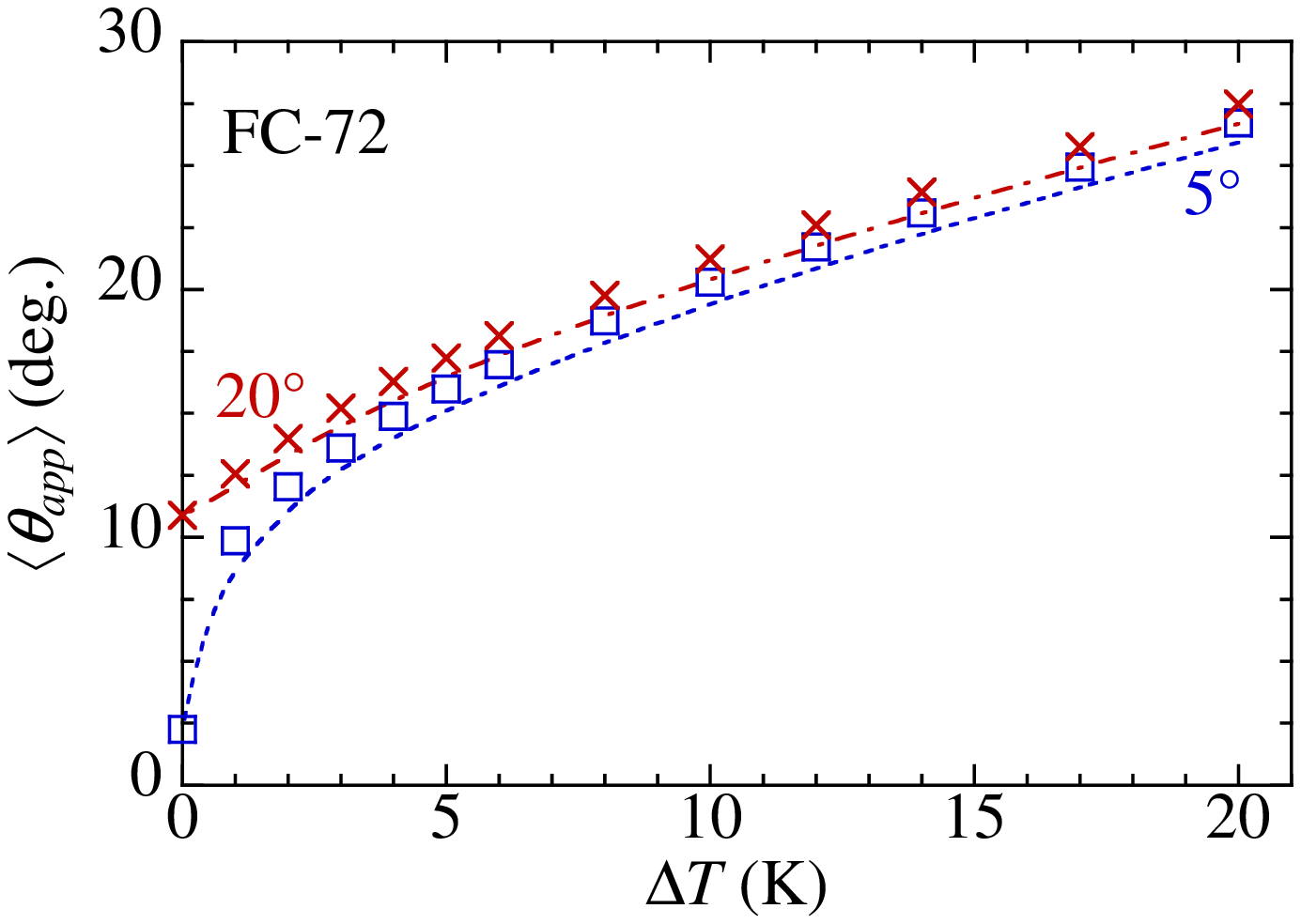}\label{fig:FC72ThetaV-LV-DT}}\\
\subfloat[]{\includegraphics[width=5cm, clip]{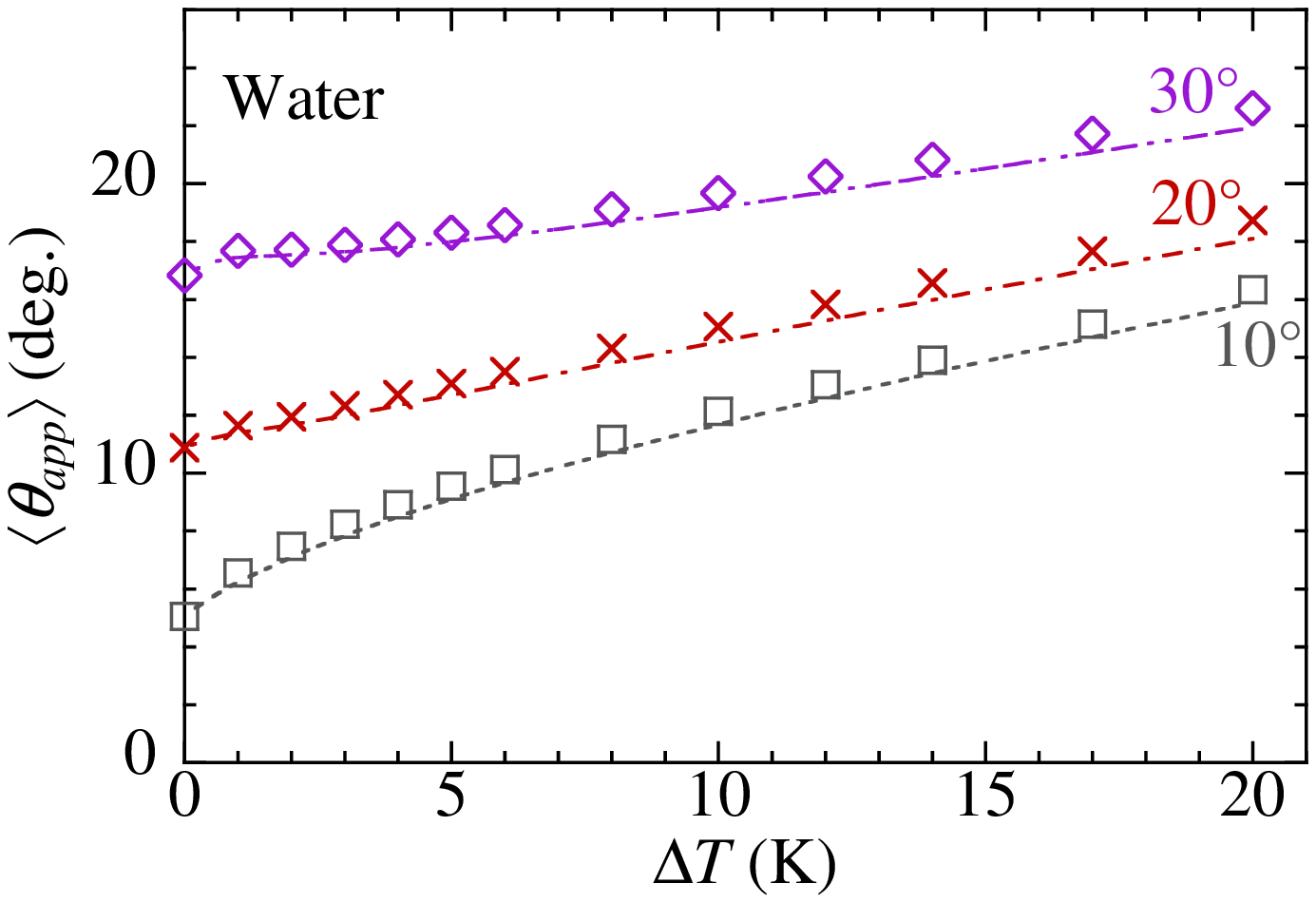}\label{fig:WaterThetaV-LV-DT}}\\
\subfloat[]{\includegraphics[width=5cm, clip]{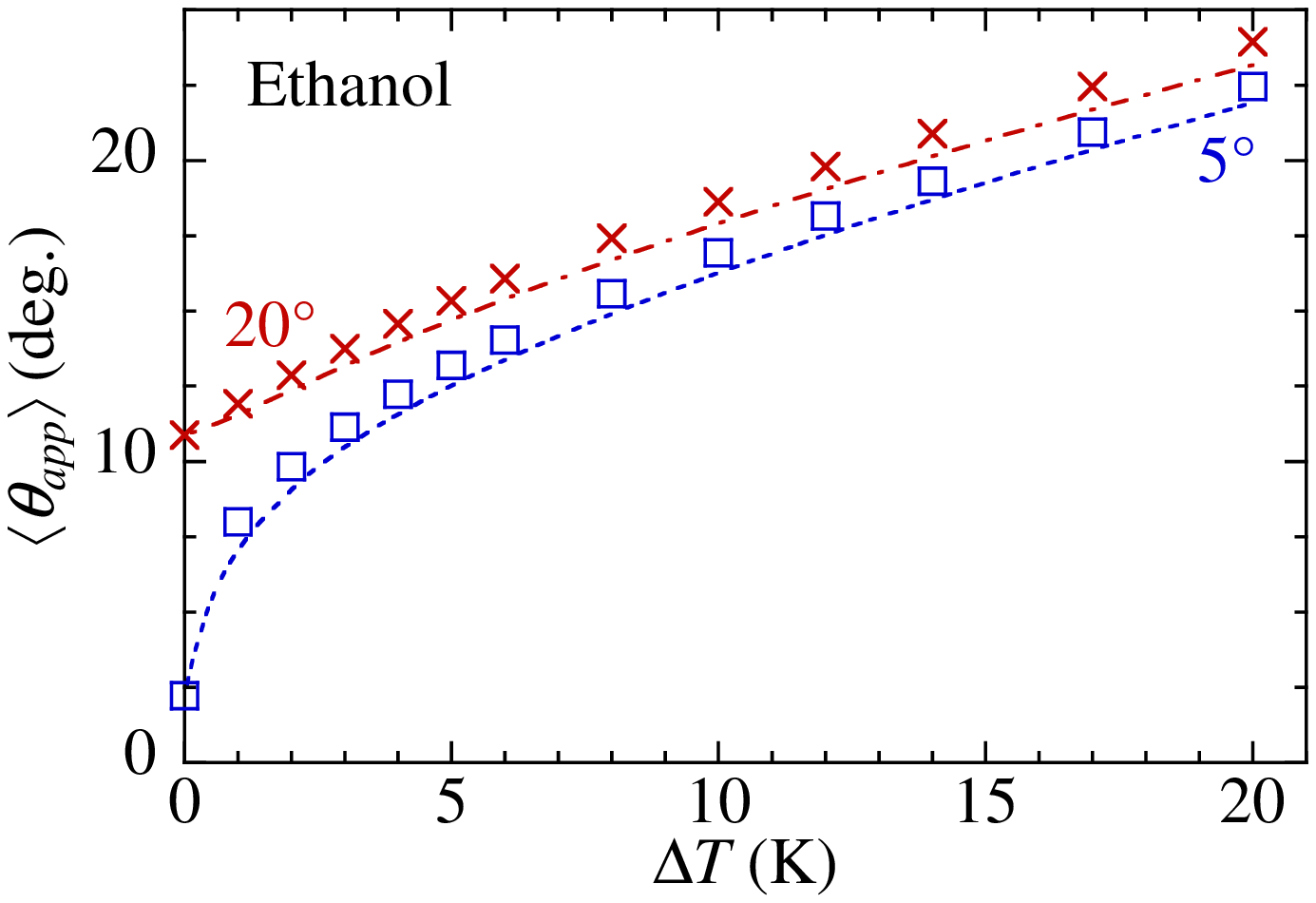}\label{fig:EthanolThetaV-LV-DT}}
\caption{Comparison of numerical (characters) and time-averaged model (lines) results of $\langle\theta_{app}\rangle (\Delta T)$ for several $\theta_{micro}$ and different fluids.}\label{fig:ThetaV-LV-DT}
\end{figure}

In applications (e.g. in pulsating heat pipe modeling), the focus is primarily on $Ca$ and $\theta_{app}$, which experience slow variations during the dewetting process. Therefore their time-averaged values are sufficient. We perform the averaging of the lubrication 2D approach \cite{JFM22} over a time interval $\Delta t=t_2-t_1$ with $t_1=0.1t_d$ and $t_2=0.6t_d$; the time-averaged quantities are denoted hereafter by the angle brackets. Such a time interval is reasonably large, as it corresponds to a half of the thin film lifetime $t_d$. Fig.~\ref{fig:CaDT} plots the $\langle Ca\rangle$ data from the lubrication approach as characters with bars for three fluids and several $\theta_{micro}$ values. The upper and lower limits of the bars correspond to the $Ca$ values at $t_1$ and $t_2$, respectively. The bars are relatively short with respect to the absolute values of $Ca$, indicating that the averaging is reasonable. Noticeably, these numerical data demonstrate that the dewetting accelerates with superheating, i.e. with the evaporation rate. As shown in fig.~\ref{fig:ThetaV-LV-DT}, $\langle\theta_{app}\rangle$ also grows with superheating.

The characters in figs.~\ref{fig:CaDT}, \ref{fig:ThetaV-LV-DT} result from the full numerical solution of eq.~\eqref{eq:GEA}. However, such a solution is time- and resource-consuming because of several reasons: (i) necessity to resolve a big range of the length scales; (ii) numerical iteration at each time moment to manage the nonlinear terms (iii) solving of a transient ridge evolution. A simpler approach is thus desirable.

The adiabatic asymptotic expressions \eqref{asymp} can be generalized \cite{JFM22} to the case $\Delta T\neq 0$ thanks to the scale separation between the three asymptotic regions discussed above. We propose here the time averaging of the above expressions. When this process is slow with respect to the dewetting dynamics, the evolution is quasi-static, and eqs.~\eqref{asymp} should hold for the time-averaged quantities:
\begin{equation}\label{asympa}
\begin{split}
  & \langle \theta_{app}\rangle^3 =\theta^3_V-9\langle Ca\rangle \ln \frac{2\langle w\rangle}{e\ell_{V}}, \\
  & \langle Ca\rangle=\frac{\theta^3_V}{9}\left[ \ln \left(  \frac{4a}{e^2} \langle Ca\rangle^{1/3} \frac{\langle w\rangle^2}{\ell_{V}\langle h_\infty\rangle } \right) \right] ^{-1}.
\end{split}
\end{equation}
Consider now all the parameters in these expressions. For given $\theta_{micro}$, $\theta_V$ can be calculated from the microregion model (fig.~\ref{fig:thetaV_DT}). As shown in \cite{JFM22}, $\ell_V$ can be approximated by the formula
\begin{equation}\label{lVslip}
\ell_{V}\simeq{3l_s}/({e\theta_V}).
\end{equation}
This expression is inspired by the adiabatic result \cite{Snoeijer10} and signifies that $\ell_{V}$ is largely controlled by the hydrodynamic slip. Two remaining parameters, $\langle w\rangle$ and $\langle h_\infty\rangle$ are yet to be defined.

During the film evaporation, $h_\infty$ decreases, which affects the dynamics. The value of $\langle h_\infty\rangle$ can be computed with eq.~\eqref{eq:hfe},
\begin{equation}
   \langle h_\infty\rangle=\frac{1}{\Delta t}\int_{t_1}^{t_2} h_0\sqrt{1-\frac{t}{t_d}} dt= \beta h_0,
\end{equation}
where the time averaging is performed over $\Delta t=t_2-t_1$; for the same time interval $\Delta t=0.5t_d$ as in fig.~\ref{fig:CaDT}, $\beta\simeq 0.8$.

Following the theory simplified in the spirit of \cite{BW}, we can calculate $\langle w\rangle$. Recall the dewetting physics; as the contact line recedes, the liquid previously covering the solid surface accumulates in the ridge (fig.~\ref{fig:DewettingRidge}). Because of capillary action, the interface profile of the dewetting ridge is approximated by a circular arc of radius $R$, and the center of the circle is marked as point O. The arc intersects the solid surface at point A of apparent contact angle $\theta_{app}$ and the liquid film surface with an angle $\varphi$. Point B on the substrate is under the ridge peak. The radius $R$ satisfies
\begin{equation}\label{eq:RidgeR}
  R\sin\theta=w,
\end{equation}
where $w=\overline{\mathrm{\mathrm{AB}}}$. The angles $\theta_{app}$ and $\varphi$ satisfy the geometrical relation $\overline{\mathrm{OC}}- \overline{\mathrm{OB}}=\overline{\mathrm{CB}}$, resulting in
\begin{equation}\label{eq:theta_phi}
  R(\cos\varphi-\cos\theta_{app})=h_{\infty}.
\end{equation}

The substitution of \eqref{eq:RidgeR} into \eqref{eq:theta_phi} and averaging leads to
%
\begin{equation}\label{eq:theta_phi2a}
\cos\langle\varphi\rangle=\cos\langle\theta_{app}\rangle + \frac{1}{\tilde w}\sin\langle\theta_{app}\rangle,
\end{equation}
where $\tilde{w} =\langle w\rangle/h_\infty$.

The volume of liquid (per unit CL length)
\begin{equation}\label{eq:S1_S2}
  S= \int_{t_1}^{t_2} h_\infty(t) U_{cl}(t) dt
\end{equation}
collected from A$'$ to A (during the time $\Delta t$) is approximately equal to the excessive ridge area
\begin{equation}\label{eq:S2}
  S=R^2(\varphi-\sin\varphi\cos\varphi),
\end{equation}
colored in light blue in fig.~\ref{fig:DewettingRidge}. Since the evolution is slow, one can replace eq.~\eqref{eq:S1_S2} with $\langle S\rangle\simeq\langle U_{cl}\rangle\langle h_\infty\rangle\Delta t$.

By combining eqs.~(\ref{eq:RidgeR}, \ref{eq:theta_phi2a}, \ref{eq:S1_S2}) and averaged eq.~\eqref{eq:S2}, we arrive at
\begin{equation}\label{eq:Ucl_geo}
  \sin^2 \langle\theta_{app}\rangle \Delta \tilde t=N {\tilde w^2}(\langle\varphi\rangle-\sin\langle\varphi\rangle\cos\langle\varphi\rangle),
\end{equation}
where $\Delta \tilde t=\Delta t/t_d$ and the dimensionless parameter
\begin{equation}\label{eq:N}
N=\langle h_\infty\rangle/(t_d \langle U_{cl}\rangle)\end{equation}
is a ratio of drying and CL receding speeds. The $N$ smallness is a necessary condition for validity of the quasi-steady approximation and then of the whole model. Another quasi-steadiness criterion is a smallness of the viscous relaxation time with respect to $t_d$, i.e. of $M=\mu\langle h_\infty\rangle/(\sigma t_d)\equiv N\langle Ca\rangle$. However, as $\langle Ca\rangle\ll 1$, this criterion always holds if $N\ll 1$ holds.

The set of algebraic equations (\ref{asympa}, \ref{lVslip}, \ref{eq:theta_phi2a}, \ref{eq:Ucl_geo}) is complete provided that $\theta_V$ is known for the working fluid and for given $\theta_{micro}$ and $\Delta T$ (fig.~\ref{fig:thetaV_DT}). We can now compute four quantities $\langle\varphi\rangle$, $\tilde{w}$, $\langle\theta_{app}\rangle$ and $\langle Ca\rangle$, the latter two being the most significant. Note that their dependence on a particular choice of $\Delta t$ is extremely weak.

The $\Delta T$ variations of $\langle\theta_{app}\rangle$ and $\langle
Ca\rangle$ produced by the time-averaged method
(lines) are compared to the numerical solution of eq.~\eqref{eq:GEA} in figs.~\ref{fig:CaDT} and \ref{fig:ThetaV-LV-DT} for several contact angles and three different fluids commonly used in heat transfer applications. A good agreement between the lubrication solution and the time-averaged theory is achieved.
It is explained by $N\sim 10^{-3}$ in all the cases considered here. Note that $N$ given by eq.~\eqref{eq:N} remains bounded at increasing $\Delta T$ due to $\langle\theta_{app}\rangle$ growth with $\Delta T$ slightly faster than linear (figs.~\ref{fig:CaDT}) thus compensating the $t_d$ decrease, cf. eq.~\eqref{td}. While remaining small, $M$ grows with $\Delta T$ nearly linearly, which probably explains a small but growing discrepancy between the lines and the characters in figs.~\ref{fig:CaDT} and \ref{fig:ThetaV-LV-DT}.

While figs.~\ref{fig:CaDT} and \ref{fig:ThetaV-LV-DT} may look similar to figs.~6 and 7 of \cite{JFM22}, they are not the same at all. Certainly, the characters representing the numerical data, which gives a reference for comparison, are the same in both cases. However, the lines are different. To obtain the lines in \cite{JFM22}, one averaged the time-dependent results for $w$, $h_\infty$, and $\ell_V$ of numerical simulations; by using these values, $\langle Ca\rangle$ was calculated with eqs.~\eqref{asymp}.  In contrast, the lines in the present Letter are produced directly by solving equations (\ref{asympa}, \ref{lVslip}, \ref{eq:theta_phi2a}, \ref{eq:Ucl_geo}), which are independent of the non-stationary lubrication simulation. Only the microregion solution is needed for them. Such an approach requires only seconds of computation time instead of weeks to obtain the results in \cite{JFM22}.

\section{Conclusions}

The dewetting phenomenon in the presence of evaporation can be described in two dimensions with the generalized lubrication approach. Its numerical solution has shown that the dewetting speed increases with the substrate superheating, indicating that evaporation accelerates the dewetting. However, this numerical approach requires solving a non-linear and non-stationary partial differential equation, which is time and resource-consuming.  To obviate heavy numerical calculations, we propose a time-averaged approach, which is based on the multiscale theory and consists of several algebraic equations. The new approach employs an integration quantity as input data, the Voinov angle, which can be obtained by solving the stationary microregion problem. This value is much easier to calculate and is familiar to many researchers. We compared the time-averaged model to the full numerical solution for different wettability and three different fluids. Thanks to the slowness of the film evaporation (with respect to dewetting), the results of this simple model are in good agreement with the numerical results. This approach is original and valuable for multiple applications involving liquid film evaporation, for instance, in the modeling of heat transfer systems such as pulsating heat pipes \cite{HEFAT22-SBPHP}, and bubble growth in boiling \cite{HEFAT22-Boil}.

\section{Acknowledgments}

The present work is supported by the project TOPDESS, financed through the Microgravity Application Program by the European Space Agency. This article is also a part of the PhD thesis of X.~Zhang, co-financed by the CNES and the CEA NUMERICS program, which has received funding from the European Union's Horizon 2020 research and innovation program under grant agreement No 800945 NUMERICS H2020-MSCA-COFUND-2017. Additional financial support of CNES awarded through GdR MFA is acknowledged.\\

\bibliographystyle{eplbib}
\bibliography{PHP,Taylor_bubbles,ContactTransf}

\end{document}